# Integrating Amdahl-like Laws and Divisible Load Theory

*Yang Cao, Fei Wu and Thomas Robertazzi*

*Stony Brook University*

**Integrating Amdahl's and Amdahl-like laws with Divisible Load Theory promises mathematical design methodology that will make possible the efficient design of tomorrow's systems.**

We are in an era where it is possible to deploy systems such as clouds/fogs/mists in technological realizations involving virtualization of computation, communication and storage.  However, we need a good base of design knowledge to do this effectively and efficiently.  As often happened over the past 70 years of the computer era, our ability to implement systems is outpacing our intellectual understanding of how they should be designed and operated. There is an urgent need to create a complete understanding to avoid the deployment of badly proportioned systems which cause economic waste, lack of accurate foresight in designing future systems and a lack of recognition of areas that call for technological improvement.

Our goal should be to make it possible to achieve a solid understanding of complex computer and information systems design using mathematical modeling.  To this end we propose integrating the foundational Amdahl's Law and variants with divisible load scheduling theory to provide such an understanding.

### Divisible Load Theory

To deal with large amount of data in modern computation system, divisible load theory (DLT) has emerged as a potential tool. Divisible loads are loads of large amounts finely parallelizable data.  The data has no precedence relationships and can be divided into parts of arbitrary size.  It is a different paradigm than atomic task scheduling.  Work since 1988 [1,2] has established means of distributing and processing such load in a time optimal fashion in many types of networks [3,4].  It is of interest when loads are in fact divisible or as an approximation in the spirit of fluid flow packet models.  Potential applications include image signal processing, big data and massive experimental data processing.

Such loads are commonly encountered in applications where a great amount of similar data units is being processed. Generally, DLT model scheduling processing occurs in two steps: load distribution and load processing.  The data is usually distributed from one (or more) processors to multiple processors and processed in parallel. An optimal schedule will be obtained to achieve the minimum finish time (makespan). Linear equations or recursions are widely used in DLT analysis, which makes it efficiently solvable.

The significance of integrating Amdahl-like laws with divisible load scheduling theory is to give designers (i.e. computer scientists and engineers) the mathematical tools to aid this growing technological revolution in much the same way as Steinmitz's mathematical work at the turn of the last century made possible over a century of systematic and tractable design of alternating current electrical systems. Today's and tomorrow's systems that will benefit from this include 5G, and systems in health care, social media, commerce, government and scientific research.

**Amdahl's Law**

Amdahl argued in 1967 [5,6] that even if one could solve the parallel part of a program in near zero time due to the use of a large number of parallel processors, the bottleneck was the sequential part of the program which could only be processed on a single processor.

The performance metric called "speedup", S, is a basic way of expressing parallel processing time advantage. It is defined as the ratio of solution time of a problem on one processor, $T(1)$, to solution time of the same problem on $n$ processors, $T(n)$:

$$S = \frac{T(1)}{T(n)} \tag{1}$$

To write this mathematically, let $f$ be the workload fraction that is parallelizable and $1 - f$ be the workload fraction that is serial. Let $n$ be the number of homogeneous (i.e. identical) processors. Let $T(1)$ be the time to solve the workload on one processor and $T(n)$ be the time to solve the workload on $n$ processors. Finally let $T_s$ be the serial execution time for the entire program.

Then:

$$T(1) = T_s, \quad T(n) = \underbrace{(1-f)T_s}_{\text{serial}} + \underbrace{f\frac{T_s}{n}}_{\text{parallel}} \tag{2}$$

Here $T(n)$ is a weighted sum of serial and parallel execution time. The parallel execution time is $fT_s/n$, the parallel workload, divided by $n$, the number of processors used. Here also it is assumed that there is no time overlap between the serial and the parallel execution.

So, one has in terms of speedup, Amdahl's Law:

$$S^{Amdahl} = \frac{T(1)}{T(n)} = \frac{1}{(1-f)+\frac{f}{n}} \tag{3}$$

In a 1988 paper J.L. Gustafson made an argument that the Amdahl Law assumption of constant problem size is usually never the case [7]. More cores are normally used to solve larger and more complicated problems. Thus, one would be justified in having a parallel fraction that grows linearly in problem size (i.e. using $fn$ instead of a single $f$). One finds [7] $S^{Gustafson} = (1 - f) + nf$. One could have a parallel fraction growth factor, $scale(n)$, that is between a constant (Amdahl's Law) and linear growth (Gustafson's Law) (see [8,9,10]). It is not the only possibility but one could use a square root function, $scale(n) = \sqrt{n}$. This leads to a general law with speedup between that of Amdahl's and Gustafson's laws.

Amdahl's Law has inspired a number of interesting and useful studies over the past years. A representative sample includes Hill and Marty who in 2008 [8,9] applied Amdahl's Law to multicore architectures and attempted to answer system level design questions. Marowka did a performance study applying Amdahl's Law to systems of CPUs and GPUs[11]. Cassidy found objective functions for average delay and average energy using Amdahl's Law [12]. Díaz-del-Río [13] presented a performance study of when it is preferable to off-load computation from a mobile device to the cloud.

**A More Complete Approach**

Over time (often closed form) expressions for divisible load model speedup have been developed for various multi-processor interconnection topology strategies and load distribution policies. Interconnection topologies include buses, stars, multi-level tree networks, meshes, hypercubes and other networks. Load distribution policies include sequential load distribution and concurrent load distribution and with simultaneous start or staggered start. In all these Amdahl's Law can be modified to be:

$$S^{Amdahl} = \frac{1}{(1-f) + \frac{f}{S_{DLT}(n)}} \qquad (4)$$

Here $S_{DLT}(n)$ is the speedup of a divisible load model of any architectured parallel facility with $n$ processors. Such a facility is a basic model that has no sequential component but considers the facility issues, which involve degrees of efficiencies due to communication delay, interconnection topology, load distribution policy and the relative difference in computation and communication intensity and speeds. Significantly these additional factors can now be included in Amdahl-like Laws.

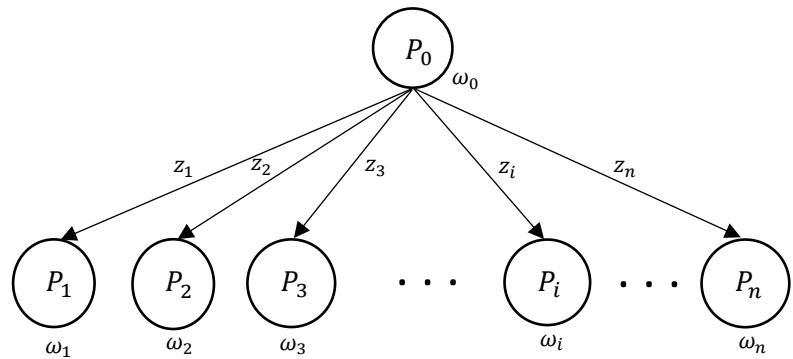

Figure 1. Single level tree network

An example of this is the single level tree network of Figure 1 where load is distributed from the root node to the children nodes. Here, $z_i$ is the $i$th link's inverse link speed and $\omega_i$ is the $i$th processor's inverse computing speed.

The boxed material indicates analytical divisible load speedup expressions for three fundamental load distribution protocols in the single level tree network (star type network) [14]. The order of processors to achieve the shortest finishing time is $z_1 \leq z_2 \leq z_3 ... \leq z_n$. The rule can be intuitively described as the processors with faster link speeds will receive load prior to the ones with slower link speeds.

The timing diagrams for communication and computation are shown in Figure 2. The first model is sequential load distribution where the source (root) node distributes load to one child processor at a time in one pass.

The second and third models involve simultaneous (concurrent) load distribution of load over all links. In the second model (staggered start) computation at a child begins only once all its computational load is received from the source node. In the third model (simultaneous start) computation at a child begins as soon as it begins to receive load.

We thus have more complete models than Amdahl's original Law.

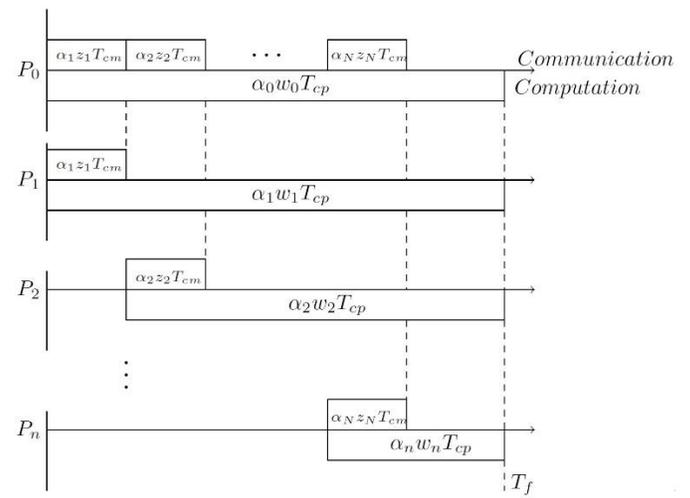

MODEL 1: Sequential Load Distribution

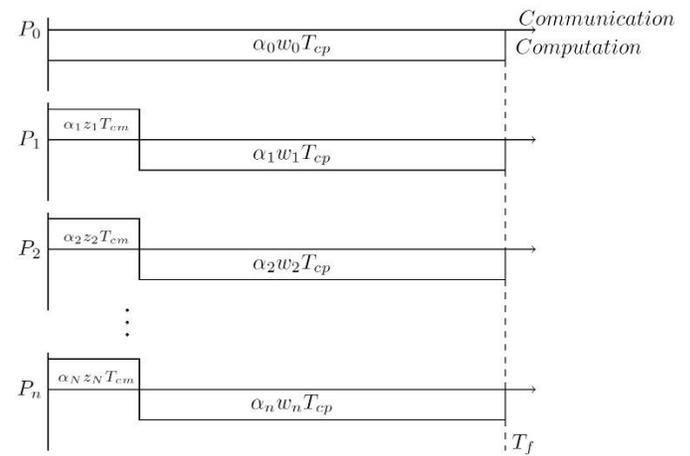

MODEL 2: Simultaneous Distribution, Staggered Start

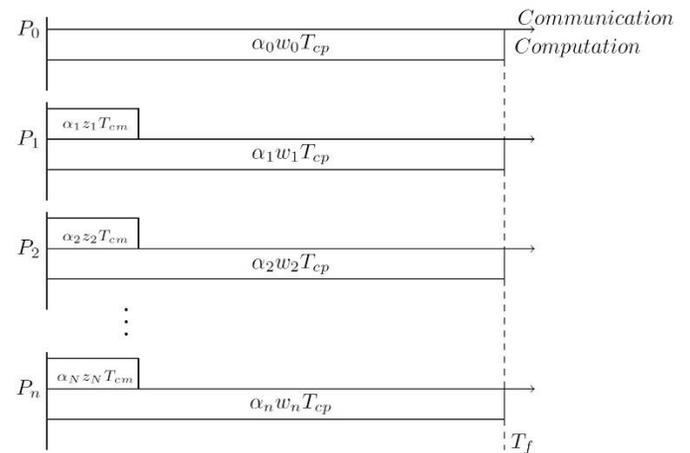

MODEL 3: Simultaneous Distribution, Simultaneous Start

Figure 2. Timing Diagrams for Three Fundamental Load Distribution Protocols

*Speedup values for different divisible load models with single level tree Networks:*

**MODEL 1: Sequential Load Distribution**

$$S_{DLT}(n) = 1 + k_1[1 + \sum_{i=2}^{n}(\prod_{l=2}^{i} q_l)] \tag{5}$$

where:

$$q_i = (\omega_{i-1}T_{cp} - z_{i-1}T_{cm})/\omega_i T_{cp}, \quad k_1 = \omega_0/\omega_1$$

For the system with homogeneous processors, the inverse processing speed and link speed of each processor (except for the root) is the same.

In this case, Equation 5 can be simplified as:

$$S_{DLT_{homo}}(n) = 1 + \frac{\omega_0}{\omega}\left[\frac{1-(1-\sigma)^n}{\sigma}\right] \tag{6}$$

where: $\quad \sigma = zT_{cm}/\omega T_{cp}$

**MODEL 2: Simultaneous Distribution, Staggered Start**

$$S_{DLT}(n) = 1 + \omega_0 T_{cp} \sum_{i=1}^{n} 1/(\omega_i T_{cp} + z_i T_{cm}) \tag{7}$$

For the system with homogeneous processors, Equation 7 can be simplified as:

$$S_{DLT_{homo}}(n) = 1 + k \times n \tag{8}$$

where: $\quad k = \omega_0 T_{cp}/(\omega T_{cp} + zT_{cm})$

**MODEL 3: Simultaneous Distribution, Simultaneous Start**

$$S_{DLT}(n) = 1 + \omega_0 \sum_{i=1}^{n}(1/\omega_i) \tag{9}$$

For the system with homogeneous processors, Equation 9 can be simplified as Equation 8, where $k = \omega_0/\omega$.

Here are the notations for the equations:

$n$: The number of processors;

$\omega_0$: The inverse of the computing speed of the source node (root node);

$\omega_i$: The inverse of the computing speed of the $i$th processor;

$z_i$: The inverse of the link speed of $i$th link;

$T_{cp}$: Computing intensity constant: the entire load is processed in $\omega_i T_{cp}$ seconds by the $i$th processor;

$T_{cm}$: Communication intensity constant: the entire load is transmitted in $z_i T_{cm}$ seconds over the $i$th link;

$S_{DLT}(n)$: The speedup with $n$ processors in the systems using a DLT model;

$S_{DLT_{homo}}(n)$: The speedup with $n$ homogeneous processors in the systems using a DLT model;

**Calculation and Analysis**

To test and compare the speedup levels for different networks, the boxed equations were inserted into Equation 4 and compared with Amdahl's original Law (Equation 3). The values used are listed in Table 1. Both systems with heterogeneous processors and homogeneous processors are tested and the results are shown in Figure 3, 4, 5 and 6. We find:

*Results Depend on Parameters:*

By comparing Figure 3 and 4, one can observe that the speedup values for the system with homogeneous processors are higher than the values of the system with heterogeneous processors for our parameters. For example, for the network topology of model 2 (with simultaneous distribution and staggered start), the speedup in Figure 3 with 30 processors is 3.86, and the speedup in Figure 4 with 30 processors is 4.25. For the same model, the curve in Figure 4 is generally higher than the one in Figure 3. This is because the processing speed for the homogeneous processors equals the highest processing speed among the heterogenous processors, which results in a higher computation power for the homogenous system.

*Simultaneous Distribution Beats Sequential Distribution:*

By comparing the values of different network topologies, one can discover that the systems with simultaneous distribution have higher speedup values than sequential distribution. This is because with simultaneous distribution, the processors can all start receiving load near the starting time, while with sequential distribution, the processors ranked lower in the sequence must wait for a considerable length of time.

*Simultaneous Start Beats Staggered Start:*

Meanwhile, the system with simultaneous start has higher speedup values than the one with staggered start. This is because staggered start means that each the processor must wait until it finishes receiving the load before starting computation. But with simultaneous start, all the processors can start computing at the time when they start to receive load.

*Amdahl's Law is an Upper Bound:*

Note that the pure Amdahl's Law prediction is the upper bound of Divisible Load Theory analysis because it does not take account divisible load based inefficiencies. This upper bound could be achieved by using model 3 and setting the root processor's computing speed to be the same as other homogeneous processors.

This upper bound is shown in both Figure 4 and Figure 6 where the system has homogeneous processors. In this case, model 3 (Simultaneous Distribution, Simultaneous Start) will have the same performance as Amdahl's Law's original analysis, which is shown in Equation 3. In our calculation, since the source processor also shares the computing task, the system in fact has $n + 1$ processors working simultaneously. So, the variable $n$ in Equation 3 is updated to be $n + 1$. We also set $\omega_0 = \omega = 4.2$. As a result, the curve of model 3 is overlapped with the Amdahl's Law Equation curve in both Figure 4 and Figure 6.

*The Influence of the Size of* the Parallelizable Load ($f$)*:*

      The speedup of different divisible load models versus different $f$ values for the systems with either heterogeneous and homogeneous processors are shown in Figure 5 and Figure 6. For the entire calculation, there are 20 children processors in the system. The parameters are the same as Table 1. Overall, when a system has a higher value of $f$ (the workload fraction that is parallelizable), it has a higher speedup. At the same time, when $f$ has a larger value, the speedup grows more quickly. For example, in Figure 6, the $S^{Amdahl}$ is around 4.2 or 6.8 for the two systems with simultaneous distribution (with either staggered start or simultaneous start) at $f = 0.8$ or $f = 0.9$. While $f = 1$, which means that all data is parallelizable, $S^{Amdahl}$ is around 20.

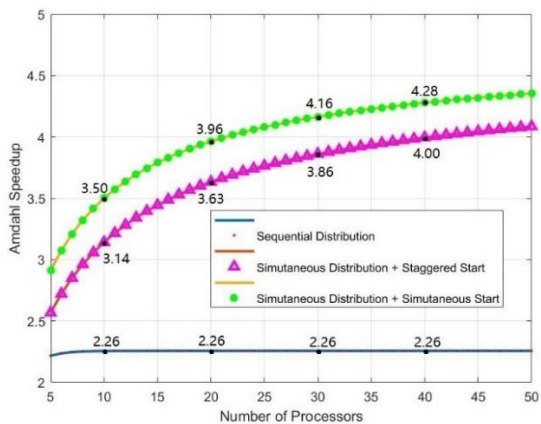

Figure 3.
Speedup of different divisible load models vs. number of heterogeneous processors

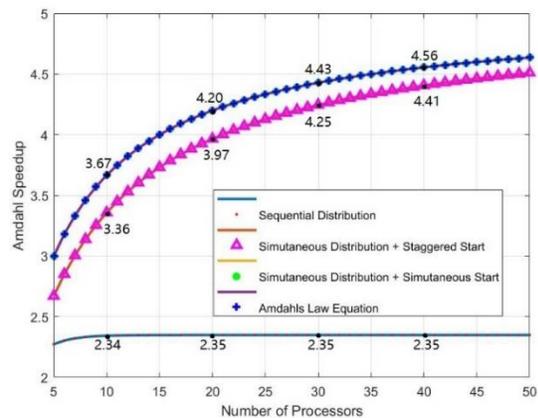

Figure 4.
Speedup of different divisible load models vs. number of homogeneous processors

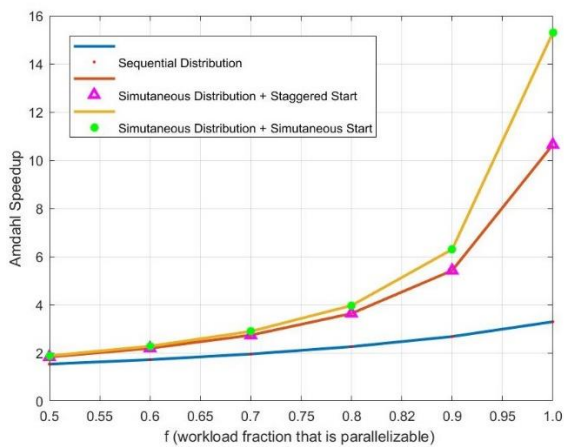

Figure 5.
Speedup of different divisible load models vs. different $f$ values for the systems with heterogeneous processors (n=20)

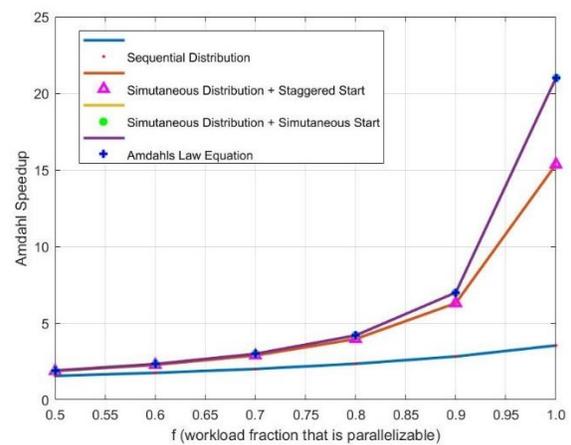

Figure 6.
Speedup of different divisible load models vs. different $f$ values for the systems with homogeneous processors (n=20)

Table 1. Values used in the calculation

| | $n$ | $\omega_0$ | $\{\omega_1, \omega_2, \omega_3, \ldots, \omega_n\}$ | $\{z_1, z_2, z_3, \ldots, z_n\}$ | $T_{cp}$ | $T_{cm}$ | f |
|---|---|---|---|---|---|---|---|
| Heterogeneous system | 50 | 4.2 | {4.2, 4.4, 4.6, …, 14} | {2.2, 2.4, 2.6, …, 12} | 2 | 1.5 | 0.8 |
| Homogeneous system | 50 | 4.2 | {4.2, 4.2, 4.2, …, 4.2} | {2.2, 2.2, 2.2, …, 2.2} | 2 | 1.5 | 0.8 |

**Significance**

The integration of Amdahl's Law and Amdahl-like laws with Divisible Load Theory is significant in showing how issues besides Amdahl's sequential/parallel paradigm may be included in an overall closed form analytical model of speedup (and makespan as well). For some more involved models $S_{DLT}(n)$ can be found numerically and inserted into Equation 4 as well. Similar integrations can be done for Gustafson's law and other Amdahl law variants. It should be mentioned that it is also possible to substitute the speedup of the pure Amdahl's law Equation 3 into the number of processors variable, $n$, in the boxed divisible load equations. The correct way to proceed would depend on the actual application.

**Conclusion**

Amdahl's law and its variations provide much to think about when evaluating the performance of parallel systems. The speedup value is a key metric while comparing the performances of different system topologies. Since parallel systems are increasingly prevalent, these issues are likely to be of interest for a considerable amount of time.

# References for
# Integrating Amdahl-like laws and Divisible Load Theory

## Yang Cao, Fei Wu and Thomas Robertazzi